\documentclass[prd,twocolumn,superscriptaddress,showpacs,amsmath,amssymb]{revtex4}
\usepackage{graphicx}
\usepackage{dcolumn}
\usepackage{bm}

\begin{document}
\title{Tachyon field in Loop Quantum Cosmology: inflation and evolution picture}

\author{Hua-Hui Xiong}
  \email{jimhard@163.com}
  \affiliation{Department of Physics, Beijing Normal University, Beijing 100875, China}

\author{Jian-Yang Zhu}
\thanks{Author to whom correspondence should be addressed}
  \email{zhujy@bnu.edu.cn}
  \affiliation{Department of Physics, Beijing Normal University, Beijing 100875, China}
\date{\today}

\begin{abstract}
Loop quantum cosmology (LQC) predicts a nonsingular evolution of the
universe through a bounce in the high energy region. We show that
this is always true in tachyon matter LQC. Different from the
classical Friedman-Robertson-Walker (FRW) cosmology, the super
inflation can appear in the tachyon matter LQC; furthermore, the
inflation can be extended to the region where classical inflation
stops. Using numerical method, we give an evolution picture of the
tachyon field with an exponential potential in the context of LQC.
It indicates that the quantum dynamical solutions have the
attractive behavior as the classical solutions do. The whole
evolution of the tachyon field is that in the distant past, the
tachyon field--being in the contracting cosmology--is accelerates to
climb up the potential hill with a negative velocity; then at the
boundary the tachyon field is bounced into an expanding universe
with positive velocity rolling down to the bottom of the potential.
In the slow roll limit, we compare the quantum inflation with the
classical case in both an analytic and numerical way.
\end{abstract}


\maketitle

\section{Introduction}

Quantum gravity is expected to rectify the classical general
relativity in the regime of high curvature where the classical
theory breaks down. Cosmology provides a stage to test this
rectification, especially in the region near the big bang
singularity. Loop quantum gravity (LQG) is a nonpertubative and
background independent approach to quantize gravity \cite {lqg}. The
underlying geometry in LQG is discrete in Planck scale. Loop quantum
cosmology (LQC) uses the framework developed in LQG to analyze the
universe \cite{lqc,mathematical-structure}. In LQC, the spatial
geometry is also discrete, and when approaching the Planck scale the
universe is described by the difference equation which can go
through the big bang point nonsingularly. Above the Planck scale the
discreteness of the spatial geometry becomes weak, and the spacetime
recovers continuum. However the dynamical equation (effective
Hamiltonian constraint) gets modification from LQC both in the
gravity and matter secter. This region is known as the semiclassical
region in LQC \cite{semiclassical}. In the semiclassical region,
based on the modification from the inverse scale factor operator,
many phenomena have been investigated, such as a natural inflation
from quantum geometry \cite{inflation-geometry}, avoidance of a big
crunch in closed cosmology \cite{avoidance-big-crunch}, appearance
of a cyclic universe \cite{cyclic-universe}, and a mass threshold of
black hole \cite {mass-threshold}, etc.

Recent investigation shows that the big bang is replaced by a big
bounce by evolving the semiclassical states backwards. And the
effective dynamics predicted by the effective Hamiltonian is shown
to match very well with the evolution of the semiclassical state
\cite{bigbang,effective-theory}. In the effective Hamiltonian
constraint, the Friedmann equation gets a quadratic density
modification for the Hubble rate $H=\dot{a}/a$, $H^2\propto \rho
\left( 1-\rho /\rho _c\right) $, which is relevant in the high
energy regime \cite{bounce}. However, this kind of modification is
independent of the inverse volume modification. Based on the
modified Friedman equation, some interesting results are obtained.
For example, it leads to generic bounces when the energy density
approaches a critical value $\rho _c$ ($\rho _c$ is about $0.82$
times the Planck density) \cite{bounce}; the scaling solutions of
the modified Friedmann equation have a dual relationship with those
in Randall-Sundrum cosmology \cite{dual}; the future singularity can
be avoided with the modified Friedmann equation \cite{future}.

In cosmology, the tachyon field might be responsible for the
cosmological inflation at early epochs and could contribute to some
new form of cosmological dark matter at late times \cite{Sen1}. In
the framework of LQC, based on the inverse volume modification in
the matter part, the author of Ref. \cite{AAAsen} has discussed the
properties of the tachyon matter field. It is shown that there
exists a super accelerated phase in the semiclassical region, and
the solution of the modified Friedmann equation corresponds to the
power law solution of the classical equation of the tachyon matter
in the classical FRW cosmology. However, as shown in Ref.
\cite{dual}, the quadratic density modification dominates over the
inverse volume modification, and the latter can be suitably
neglected when the value of the half-integer parameter $j$ (which
marked the inverse volume modification) is small. In Ref.
\cite{Preprration}, we discuss the role of $j$ for the effective
dynamics, and it is shown that neglecting the inverse volume
modification does not affect the behavior of the modified Friedmann
equation qualitatively. Therefore, in this paper we shall
investigate the behavior of the tachyon matter field in the context
of LQC based on the $\rho ^2$ modification and neglect the inverse
volume modification. It is expected that LQC will greatly change the
behavior of the tachyon field in the high energy region, especially
when the critical density $\rho _c$ is approached. We find that, for
the tachyon matter cosmology, there exists the superaccelerated
phase in the region $\frac 12\rho _c<\rho <\rho _c$ as the usual
scalar field in LQC \cite{dual}, and the inflationary e-folding can
be increased due to the LQC modification. It is difficult to obtain
the exact solutions for the modified Friedman equation and the
Hamiltonian equations for the tachyon matter, therefore we analyze
the evolution of the tachyon field in LQC by the numerical method.
The numerical results show that the evolution of the tachyon matter
LQC behaves very differently than in FRW cosmology.

In this paper we only treat the tachyon field as a scalar field with
a nonstandard kinetic term, without claiming any identification of
the tachyon field with the string tachyon field.

This paper is organized as follows. In Sec. \ref{Sec. 2}, we will
introduce the tachyon field into the effective theory of LQC and
present the state parameter equation in the modified Friedmann
equation. Then in Sec. \ref {Sec. 3}, we use the numerical method to
study the evolution of the tachyon field with an exponential
potential in LQC, and the slow roll inflation is also investigated.
Finally, Sec. \ref{Sec. 4} is the conclusion.

\section{Tachyon matter in loop quantum cosmology}

\label{Sec. 2}In this section, we will first introduce the tachyon
field in the FRW cosmology, then the Hamiltonian formulation of the
tachyon field can be put in the context of LQC where the dynamics is
described by the effective Hamiltonian.

According to Sen \cite{Sen1}, in a spatially flat FRW cosmology the
Hamiltonian for the tachyon field can be written as
\begin{equation}
H_\phi \left( \phi ,\Pi _\phi \right) =a^3\sqrt{V^2\left( \phi
\right) +a^{-6}\Pi _\phi ^2},  \label{a1}
\end{equation}
where $\Pi _\phi $ is the conjugate momentum for the tachyon field $\phi $, $%
V\left( \phi \right) $ is the potential term for the tachyon field,
and $a$ is the FRW scale factor. Here, we start with the Hamiltonian
formulation but not the Lagrangian form, because LQC takes the
canonical quantization procedure where the dynamical law is
described by the Hamiltonian constraint.

Now the Hamiltonian equations are given by
\begin{equation}
\dot{\phi}=\frac{\partial H_\phi }{\partial \Pi _\phi
}=\frac{a^{-3}\Pi _\phi }{\sqrt{V^2+a^{-6}\Pi _\phi ^2}},
\label{a2}
\end{equation}
\begin{equation}
\dot{\Pi}_\phi =-\frac{\partial H_\phi }{\partial \phi }=-\frac{%
a^3VV^{^{\prime }}}{\sqrt{V^2+a^{-6}\Pi _\phi ^2}},  \label{a3}
\end{equation}
where the prime denotes $dV\left( \phi \right) /d\phi $. From Eq. (\ref{a2}%
), one can also obtain the conjugate momentum as
\begin{equation}
\Pi _\phi =\frac{a^3V\dot{\phi}}{\sqrt{1-\dot{\phi}^2}}  \label{a4}
\end{equation}
which is just the definition of the conjugate momentum for the
tachyon field in the Lagrangian formulation.

Using the Hamiltonian equations, one can get the evolution equation
of the tachyon field as
\begin{equation}
\ddot{\phi}=-\left( 1-\dot{\phi}^2\right) \left( 3H\dot{\phi}+\frac{%
V^{^{\prime }}}V\right)  \label{a5}
\end{equation}
where $H=\dot{a}/a$ is the Hubble parameter.

In the Hamiltonian formulation, one cannot identify the matter
density and pressure for the tachyon field in the usual way by
varying the action of the tachyon field with respect to the
spacetime metric. However, one can define the energy density and the
pressure as \cite{pressure}
\begin{equation}
\rho _\phi =a^{-3}H_\phi ,\ P_\phi =-\frac 13a^{-2}\frac{\partial H_\phi }{%
\partial a}.  \label{a6}
\end{equation}

Using Eq. (\ref{a4}), one can find that the expressions of the
matter density and pressure obtained from the above definition are
the same as those in the Lagrangian formulation:
\begin{equation}
\rho _\phi =\frac V{\sqrt{1-\stackrel{.}{\phi }^2}},P_\phi =-V\sqrt{1-%
\stackrel{.}{\phi }^2}.  \label{a7}
\end{equation}

Next, we will introduce the tachyon field in the context of LQC, we
will focus on the semiclassical region of LQC where the evolution of
the universe is described by the effective Hamiltonian constraint
and the spacetime recovers the continuum.

LQG is a canonical quantization of gravity. In LQG, the phase space
of the
classical general relativity is expressed in terms of SU(2) connection $%
A_a^i $ and densitized triads $E_i^a$. In the homogeneous and
isotropic cosmology, the symmetry of spacetime reduces the phase
space of infinite freedom degrees to finite ones. Thus, in LQC the
classical phase space consists of the conjugate variables of the
connection $c$ and triad $p$, which satisfy Poisson bracket $\left\{
c,p\right\} =\frac 13\gamma \kappa $, where $\kappa =8\pi G$ ($G$ is
the gravitational constant) and $\gamma $ is the Barbero-Immirzi
parameter which is fixed to be $\gamma \approx 0.2375$ by the black
hole thermodynamics. For the flat model, the new variables have the
relations with the metric components of the FRW cosmology as
\begin{equation}
c=\gamma \dot{a},\quad p=a^2  \label{a8}
\end{equation}
where $a$ is the FRW scale factor. In terms of the connection and
triad, the classical Hamiltonian constraint is given by
\cite{mathematical-structure}
\begin{equation}
H_{cl}=-\frac 3{\kappa \gamma ^2}\sqrt{p}c^2+H_M.  \label{a9}
\end{equation}

For quantization in LQC, the elementary variables are the triads and
holonomies of connection along an edge which is defined as
$h_i\left( \mu \right) =\cos \left( \mu c/2\right) +2\sin \left( \mu
c/2\right) \tau _i$, where $\mu $ is the length of the $i$th edge
with respect to the fiducial
metric, and $\tau _i$ is related to Pauli matrices as $\tau _i=-\frac i2%
\sigma _i$. There is no quantum operator corresponding to the connection $c$%
, but the holonomies and the triads have well defined quantum
operators such that for quantization the Hamiltonian constraint must
be reformulated as the elementary variables, i.e., the holonomies
and triad. So, the Hamiltonian constraint operator can be obtained
by promoting the holonomies and triad to the corresponding
operators. The underlying geometry in LQC\ is discrete, and the
quantum Hamiltonian constraint incorporating this discreteness can
evolve through the big bang point nonsingularly.

Some long standing questions remain to be systematically answered,
such as whether the cosmology can evolve through the classical
singular point, etc. The semiclassical state is constructed in
\cite{bigbang}, and evolving the semiclassical state backward in the
expanding universe reveals that near the singularity the universe
bounces into a contracting branch. It also shows that the quantum
feature of the universe can be well described by the effective
theory which predicted a quadratic matter density correction for the
modified Friedmann equation. For the effective theory, the region is
above the Planck scale where the spacetime recovers the continuum,
and the dynamical equation takes the usual differential form.

The effective Hamiltonian constraint is given by \cite
{bigbang,effective-theory}
\begin{equation}
H_{eff}=-\frac 3{\kappa \gamma ^2\bar{\mu}^2}\sqrt{p}\sin ^2\left( \bar{\mu}%
c\right) +H_M,  \label{a10}
\end{equation}
where $\bar{\mu}$ is related to the physical area of the square loop
over which the holonomies are computed, and the area is
$\bar{\mu}^2p=\alpha \ell
_{Pl}^2$ ($\alpha $ is order one and $\ell _{Pl}=\sqrt{G\hbar }$, where $%
\hbar $ is the reduced Planck constant) fixed by the minimal area in
LQG. Here, the matter Hamiltonian $H_M$ is expressed by the tachyon
field $H_\phi $ given by the equation (\ref{a1}). Using the
Hamiltonian constraint (\ref {a10}) one can get the Hamiltonian
equation
\begin{equation}
\dot{p}=\left\{ p,H_{eff}\right\} =-\frac{\gamma \kappa
}3\frac{\partial H_{eff}}{\partial c}.  \label{b1}
\end{equation}
Squaring the above equation and substitution of the vanishing
Hamiltonian constraint $H_{eff}\approx 0$, the modified Friedmann
equation can be obtained as
\begin{equation}
H^2=\frac \kappa 3\rho _\phi \left( 1-\frac{\rho _\phi }{\rho
_c}\right) , \label{b2}
\end{equation}
where $\rho _c=\frac 3{\kappa \gamma ^2\bar{\mu}^2a^2}=\frac
3{\kappa \gamma ^2\alpha \ell _{Pl}^2}$. Here, the matter density
follows the definition in the Eq. (\ref{a7}). As analyzed in
\cite{bounce}, the modified Friedmann equation predicts a
nonsingular bounce when the matter density approaches the critical
value $\rho _c$.

In the context of the effective Hamiltonian, the Hamiltonian
equations of the tachyon field behave as its classical ones, so the
energy conservative equation is still satisfied as
\begin{equation}
\dot{\rho}_\phi +3H\left( \rho _\phi +P_\phi \right) =0.  \label{b3}
\end{equation}
Differentiating the modified Friedmann equation with respect to
time, then using Eq. (\ref{b3}) above, we get
\begin{eqnarray}
2\dot{H} &=&-\kappa \left( \rho _\phi +P_\phi \right) \left(
1-\frac{2\rho
_\phi }{\rho _c}\right)  \nonumber \\
&=&-\kappa \frac V{\sqrt{1-\dot{\phi}^2}}\left( 1-\frac 1{\rho _c}\frac{2V}{%
\sqrt{1-\dot{\phi}^2}}\right) .  \label{b4}
\end{eqnarray}
From the above equation we know that when the energy density is in
the range: $\frac 12\rho _c<\rho _\phi <\rho _c$, $\dot{H}$ is
always greater than zero. This implies that there exists a
superinflation phase for the tachyon matter cosmology. Here, the
quantum geometry modification is incorporated, which is different
from the inverse volume modification where the superinflation
depends on the quantization ambiguity parameter $j$ \cite {AAAsen}.

As in Ref. \cite{bounce}, we can analyze the bounce behavior of the
tachyon matter cosmology. Whether a bounce or recollapse happens
depends on the sign of $\ddot{p}\left| _{\dot{p}=0}\right. $.
Differentiating the Eq. (\ref{b1}) with respect to the time and
using the Hamiltonian equation about $\dot{c}$ given by the
Hamiltonian constraint (\ref{a10}), one can get
\begin{equation}
\ddot{a}\left| _{\dot{a}=0}\right. =\frac \kappa 2a\left( \rho _c-\frac 13%
a^{-2}\frac{\partial H_\phi }{\partial a}\right) =\frac \kappa
2a\left( \rho _c+P_\phi \right)   \label{b5}
\end{equation}
where the second line uses the definition of the pressure. For the
tachyon
field the state parameter equation is $\omega =\frac{P_\phi }{\rho _\phi }%
=-\left( 1-\dot{\phi}^2\right) $, so the above equation can be
rewritten as
\begin{equation}
\ddot{a}\left| _{\dot{a}=0}\right. =\frac \kappa 2a\rho _c\left(
1+\omega \right) =\frac \kappa 2a\rho _c\dot{\phi}^2\geqslant 0.
\label{b6}
\end{equation}
Therefore, in the context of LQC for the tachyon matter cosmology, a
bounce always exists by which a singularity can be avoided.

Now, we can obtain the modified Raychaudhuri equation as
\begin{eqnarray}
\frac{\ddot{a}}a &=&\dot{H}+H^2=-\frac \kappa 6\left\{ \rho _\phi \left( 1-%
\frac{\rho _\phi }{\rho _c}\right) \right.   \nonumber \\
&&\left. +3\left[ P_\phi \left( 1-\frac{\rho _\phi }{\rho _c}\right) -\frac{%
\rho _\phi ^2}{\rho _c}\right] \right\} .  \label{b7}
\end{eqnarray}
Comparing with the classical Friedmann and Raychaudhuri equation,
the effective matter density and pressure can be identified as
\begin{equation}
\rho _{eff}=\rho _\phi \left( 1-\frac{\rho _\phi }{\rho _c}\right) ,
\label{b8}
\end{equation}
\begin{equation}
P_{eff}=P_\phi \left( 1-\frac{\rho _\phi }{\rho _c}\right)
-\frac{\rho _\phi ^2}{\rho _c},
\end{equation}
then the modified Raychaudhuri equation can be rewritten as
\begin{equation}
\frac{\ddot{a}}a=-\frac \kappa 6\left( \rho _{eff}+3P_{eff}\right) .
\label{b9}
\end{equation}
One can easily checked that the effective matter density and the
pressure still satisfies the conservative equation
\begin{equation}
\dot{\rho}_{eff}+3H\left( \rho _{eff}+P_{eff}\right) =0.
\label{b10}
\end{equation}

For $\rho _\phi \ll \rho _c$, one can get $\rho _{eff}\sim \rho _\phi $ and $%
P_{eff}\sim P_\phi $, then
\[
\frac{\ddot{a}}a=-\frac \kappa 6\left( \rho _\phi +3P_\phi \right) .
\]
It means that in the case of $\rho _\phi \ll \rho _c$ the
Raychaudhuri equation behave as classically as the modified
Friedmann equation and the quantum effect can be neglected.

For $\rho _\phi \rightarrow \rho _c$ (high energy density region,
the critical density $\rho _c$ is on the order of Planck density)
the modification from the quantum geometry dominates, such that the
classical singularity is replace by a bounce. At at the bounce point
$\rho _{eff}\sim 0 $ and $P_{eff}\sim -\rho _c$, so a superinflation
happens after the bounce. In the superinflation period, the
effective matter density increase with the expansion of the universe
until the effective matter density attains the maximum value $\frac
12\rho _c$, then the superinflation phase ends. In the classical
tachyon matter cosmology the evolution of the tachyon field can not
cross the $\omega =-1$, so the superinflation of tachyon matter in
LQC purely comes from the effect of the quantum geometry.

Furthermore, for the modified Raychaudhuri equation, the inflation
stops when the right hand side of the Eq. (\ref{b9}) satisfies $\rho
_{eff}+3P_{eff}=0$. So at the inflection point, the matter density
is
\begin{equation}
\rho _\phi =\frac{3\omega +1}{3\omega +4}\rho _c.  \label{c1}
\end{equation}
For the tachyon matter field, $\omega \in \left[ -1,0\right] $. If
the state parameter $\omega =-\frac 13$ at the inflection point,
then $\rho _\phi =0$. This means that for the case of $\omega
=-\frac 13$ the end of the inflation needs matter density $\rho
_\phi =0$. However, really for tachyon matter field, $\omega =-\frac
13\Rightarrow $ $\rho _\phi =-3P_\phi >0$. So, for the tachyon
matter field , classically the inflation ends when the state
parameter $\omega =-\frac 13$; however, in the context of LQC, the
inflation phase can extend to the region where the classical
inflation stops. The effect of the extended inflation phase is only
notable when the inflation ends at the high energy region. If the
inflation happens at the low energy region (i.e., $\rho _\phi \ll
\rho _c$) where the quantum effect can be neglected, the quantum and
classical inflation approaches the same trajectory.

\section{Evolution of tachyon field with an exponential potential in LQC}

\label{Sec. 3}

\subsection{Evolution picture}

In this subsection, we will investigate the evolution of the tachyon
field in the context of LQC. It is difficult to obtain the analytic
solution for the modified Friedmann equation, so we will use
numerical method to tackle this issue. In the space of solutions for
the tachyon field with an exponential potential, the inflation
attractor has been discussed \cite {attractor}. From the above
analysis we know that the inflation of the tachyon field is
independent of any initial condition, and it is purely coming from
the quantum geometry effect. It is expected that the inflation
attractor still can be kept for the effective dynamics. Using the
numerical analysis of the evolution of the tachyon field, we find
that it is true and different from the classical FRW cosmology. The
inflationary attractor appears both in the expanding and contracting
cosmology in the context of LQC. In the following, we will discuss
these results.

Because of the Hamiltonian constraint, the four dimensional
dynamical phase
space ($c\left( t\right) ,p\left( t\right) ,\phi \left( t\right) ,\dot{\phi}%
\left( t\right) $) reduces to three dimensional one. Here, we will
show the two dimensional phase portrait consisting of $\phi $ and
$\dot{\phi}$. As shown in Ref. \cite{bounce}, for the evolution of
the tachyon matter cosmology the bounce is a critical point which
connects the expanding and contracting branches. The cosmology
evolving forward at the bounce instant enters the expanding branch.
Conversely it is in the contracting branch.

Now, substituting the modified Friedmann Eq. (\ref{b2}) into the
conservative Eq. (\ref{b3}), and using the expressions of the energy
density and the pressure (\ref{a7}), one can obtain the dynamical
equation of the tachyon matter field $\phi $
\begin{eqnarray}
\ddot{\phi} &=&-\left( 1-\dot{\phi}^2\right) \frac{V^{^{\prime }}}V\mp 3\dot{%
\phi}\left( 1-\dot{\phi}^2\right)   \nonumber \\
&&\times \left[ \frac \kappa 3\frac V{\sqrt{1-\dot{\phi}^2}}\left( 1-\frac V{%
\rho _c\sqrt{1-\dot{\phi}^2}}\right) \right] ^{1/2},  \label{c2}
\end{eqnarray}
where prime denotes $\frac d{d\phi }$. Here, the ``$-$'' sign
describes the expanding branch of the tachyon field, the ``$+$''
sign expresses the contracting branch. Here the potential for the
tachyon field is taken as\cite {Sen2}
\[
V\left( \phi \right) =V_0\exp \left( -\alpha \phi \right) ,
\]
where $V_0$ is positive constant and $\alpha $ is the tachyon mass.

For the tachyon matter cosmology the bounce occurs at the matter density $%
\rho _\phi =\rho _c$, i.e.,
\begin{equation}
\frac{V_0\exp \left( -\alpha \phi \right)
}{\sqrt{1-\dot{\phi}^2}}=\rho _c. \label{c3}
\end{equation}
The critical density constraints the boundary for the evolution of
the tachyon field which lies in the band given by the Eq.
(\ref{c3}). For the
tachyon density, the different values of the constant $V_0$ and the mass $%
\alpha $ only give out the different locations of the boundary in
the phase space. The qualitative properties of the evolution can not
be changed by the different values of the parameter $V_0$ and
$\alpha $. In our numerical calculation we take the constant
$V_0=0.82$ and the mass $\alpha =0.5$. For the boundary $\rho _\phi
=\rho _c$, $\phi $ is in the region $\left[ 0,\infty \right] $.

\begin{figure}[tbp]
\includegraphics[clip,width=0.45\textwidth]{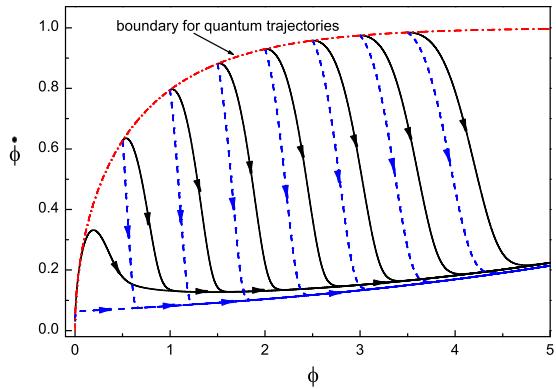}
\caption{Phase portrait for the expanding branch of tachyonic LQC.
The classical trajectories denoted by dash lines are not limmited by
the boundary; here we only show the the classical evolution
trajectoreis in the inner part of the boundary.} \label{Fig.1}
\end{figure}

\begin{figure}[tbp]
\includegraphics[clip,width=0.45\textwidth]{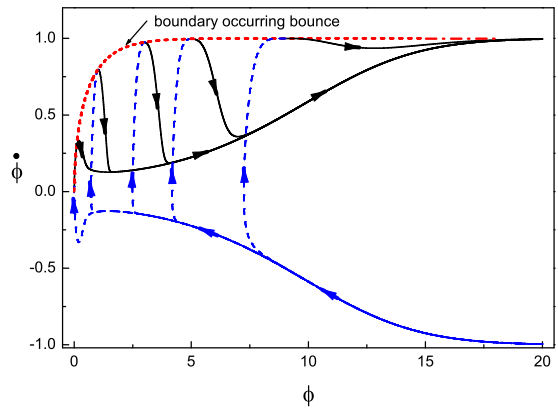}
\caption{Sketch of the complete evolution of tachyonic LQC. Here,
the solid lines and dash lines denote, respectively, the expanding
branch and the contracting branch. Also, the bounce occurs on the
boundary which connects the two branches.} \label{Fig.2}
\end{figure}

Figure \ref{Fig.1} shows the expanding branch of the tachyon
cosmology. The phase space trajectories obtained from the LQC
dynamics are shown as solid lines, and the classical trajectories
given by Eq. (\ref{a5}) are shown as
dashed lines. Figure \ref{Fig.1} indicates that for different values of $%
\phi $ and $\dot{\phi}$ on the boundary with the rolling of the
tachyon field the quantum trajectories approach the same line. This
implies that the quantum dynamics do not affect the attractor of the
solutions as do the classical solutions analyzed in Ref.
\cite{attractor}. Figure 1 also shows the attractor behaviors of the
classical solutions. With the tachyon field rolling down to larger
value of $\phi $, the quantum and classical attractors close to each
other and merge. Near the boundary, it is in the high energy region
where the quantum effect is notable, and the classical trajectories
deviate from the quantum ones. The classical trajectories can evolve
out of the boundary towards a singularity. However, on the boundary
the quantum trajectories are bounced into the contracting cosmology.
The complete evolution of the quantum trajectories is shown in the
Fig.\ref {Fig.2}.

Figure \ref{Fig.2} shows the complete evolution of the tachyon field
in LQC. In general, for an inflationary tachyon matter field with an
exponential potential, only the first quadrant of the phase space is
used, but not the others. In LQC, however, for the phase space of
the tachyon field, the first quadrant describes an expanding
cosmology with inflation in the high energy region, and the fourth
quadrant gives out a contracting cosmology which also has the
attractor behavior on the space of solutions. The bounce occurs on
the boundary which connects the two branches. Figure \ref{Fig.2}
tells us that: for the expanding branch, with the evolution of the
tachyon field, the evolution velocity $\dot{\phi}$ tends to $1$ and
the filed $\phi $ increases to $\infty $; for the contracting
branch, with the evolution of the tachyon
field to the large value, the evolution velocity $\dot{\phi}$ approaches $-1$%
. With the increasing field $\phi $, the potential of the tachyon
field monotonously decreases and the tachyon field rolls down toward
the minimum of the potential. Thus, the whole evolution of the
tachyon field can be described as the following: in the distant
past, the field, being in the contracting branch, with a negative
velocity $-1$ is accelerated climbing up the potential hill; and
then the field is bounced into an expanding universe with positive
velocity rolling down to the bottom of the potential.

\begin{figure}[tbp]
\includegraphics[clip,width=0.45\textwidth]{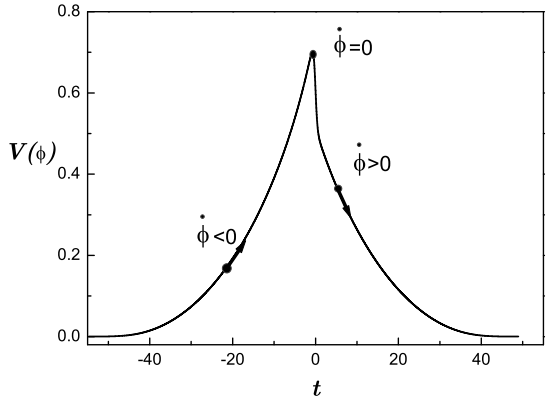}
\caption{Sketch of the potential for tachyon field with respect to
the evolution time. For different quantum trajectory the maximum of
the potential is different; here we choose the quantum trajectory
for which the field $\phi $ takes $\phi =1.0$ at the boundary to
show the potential.} \label{Fig.3}
\end{figure}

The maximum of the exponential potential is determined by the
minimal value of the field $\phi $ not corresponding to the bounce
point. In the phase space, the maximum of the potential is at the
velocity $\dot{\phi}=0$, where the field $\phi $ takes the minimum
value. This can be easily understood, since a negative $\dot{\phi}$
always decreases the value of $\phi $ leading to an increasing
exponential potential and conversely a positive velocity decreases
the potential. So the maximum potential lies at $\dot{\phi}=0$.
Figure \ref{Fig.3} shows the rolling of the tachyon field for the
potential with the parameter time $t,$ where $t=0$ is set to be the
bounce point. In LQC, the tachyon field does not monotonously rolls
down to the bottom of the potential as the classical tachyon field
dose, but it is driven to climb up the potential hill, and it then
rolls down to the minimum of the potential. The sketch of the
potential for tachyon field with respect to the evolution time is
shown in Fig.\ref{Fig.3}. Furthermore, in LQC, we can say that a
negative velocity of the field helps the tachyon field climb up the
potential hill and a positive velocity makes the tachyon rolls down
to the bottom of the potential.

\subsection{Slow roll inflation}

For the tachyonic cosmology, a successful inflation can be described
by the slow roll condition \cite{attractor}. Figure \ref{Fig.1}
shows that for the small value of $\dot{\phi}$ the quantum
trajectories approach the classical inflationary attractor. So, we
can employ the same slow roll condition to analyze the inflationary
e-folding for the tachyonic LQC.

In the slow roll limit,
\begin{equation}
\frac{\dot{a}\left( t\right) }{a\left( t\right) }\backsimeq \sqrt{\frac %
\kappa 3V\left( \phi \right) \left( 1-\frac{V\left( \phi \right) }{\rho _c}%
\right) },  \label{d1}
\end{equation}
and the Hamiltonian Eq.(\ref{a5}) becomes
\begin{equation}
3H\dot{\phi}\backsimeq -\frac{V^{^{\prime }}\left( \phi \right)
}{V\left( \phi \right) }=\alpha .  \label{d2}
\end{equation}

On the right-hand side of Eq.(\ref{d1}) the term $\left( 1-V\left(
\phi \right) /\rho _c\right) $ marks the quantum geometry
modification. In order to obtain the analytic behavior of tachyonic
LQC, we approximately replace
the term $\left( 1-V\left( \phi \right) /\rho _c\right) $ by a constant $%
\lambda $. Because $0<1-V\left( \phi \right) /\rho _c<1$, accordingly $%
0<\lambda <1$, and $\lambda =1$ corresponds to the classical case
without the quantum modification.

Now, Eq.(\ref{d1}) can be written as
\begin{equation}
\frac{\dot{a}\left( t\right) }{a\left( t\right) }\backsimeq \sqrt{\frac %
\kappa 3\lambda V\left( \phi \right) }.  \label{d3}
\end{equation}
Substituting Eq.(\ref{d3}) to Eq.(\ref{d2}) and integrating it, the field $%
\phi $ can be expressed as
\begin{equation}
\phi \left( t\right) =-\frac 2\alpha \ln \left( C-\frac{\alpha ^2}{6\gamma }%
t\right) ,  \label{d4}
\end{equation}
where $C=\exp \left( -\frac \alpha 2\phi _i\right) $ ($\phi _i$
denotes the value of $\phi $ at the start of inflation) and $\gamma
=\sqrt{\frac{\kappa V_0}3\lambda }$. Combination of the expression
$\phi \left( t\right) $ with Eq.(\ref{d3}) leads to the times of the
inflation as \cite{attractor}
\begin{equation}
\frac{a\left( t_{end}\right) }{a_i}=\exp \left[ \frac 3{\alpha
^2}\left( \lambda \frac{\kappa V_0}3C^2-\frac{\alpha ^2}6\right)
\right] ,  \label{d5}
\end{equation}
where $a_i\left( t_i=0\right) $ and $a\left( t_{end}\right) $
respectively denote the values of the scale factor at the start and
the end of the inflation.

The above analysis indicates that, for an exponential potential in
the slow roll limit, the e-folding number of the tachyonic LQC is
smaller than the classical one, i.e.,
\begin{equation}
\left. \frac{a\left( t_{end}\right) }{a_i}\right| _{quan,\ \lambda
<1}<\left. \frac{a\left( t_{end}\right) }{a_i}\right| _{cl,\text{
}\lambda =1}.  \label{d6}
\end{equation}

Using the evolution Eq.(\ref{c2}) for the tachyon matter LQC, the
numerical calculation for the quantum e-folding can be obtained as
\begin{widetext}
\begin{eqnarray*}
N_{quan}\left| _{\phi _i=0.001,\ \dot{\phi}_i=0.0316}\right.  &=&\ln \frac{%
a\left( t_{end}\right) }{a_i}=\int_{t_i}^{t_{end}}Hdt=\int_{t_i}^{t_{end}}%
\sqrt{\frac \kappa 3\rho _\phi \left( 1-\frac{\rho _\phi }{\rho _c}\right) }%
dt \\
&=&\int_{t_i}^{t_{end}}\sqrt{\frac \kappa 3\frac{V_0\exp \left(
-\alpha \phi \right) }{\sqrt{1-\dot{\phi}^2}}\left( 1-\frac{V_0\exp
\left( -\alpha \phi \right) }{\rho _c\sqrt{1-\dot{\phi}^2}}\right)
}dt\simeq 42,
\end{eqnarray*}
\end{widetext}
where the constant $V_0$ and the tachyon mass $\alpha
$ take the values as in the above subsection, $\kappa =8\pi G$
($G=1$ for the nature unit), $\phi _i$ and $\dot{\phi}_i$ are the
initial values for the inflation bounded by the boundary condition.
Here, the end time $t_{end}$ of the inflation is determined by the
condition $\omega _{eff}:=P_{eff}/\rho _{eff}=-\frac 13$. Similarly,
using the numerical method for the classical inflation the number of
the e-folding, we find that
\[
N_{cl}\left| _{\phi _i=0.001,\ \dot{\phi}_i=0.0316}\right.
=\int_{t_i}^{t_{end}}Hdt=\int_{t_i}^{t_{end}}\sqrt{\frac \kappa 3\rho _\phi }%
dt\simeq 84,
\]
where $\phi _i$ and $\dot{\phi}_i$ are the initial values for the
inflation and they take values on the boundary. It is clear that the
numerical calculations shows the result given by the Eq.(\ref{d6}).

A successful inflation needs sixty e-folding. From the
Eq.(\ref{d5}), we know that for tachyon matter LQC a sufficient
inflation favors a bigger constant $V_0$ in the exponential
potential or smaller tachyon mass $\alpha $ compared with the values
of the parameters used in our numerical calculation. For a small
mass $\alpha =0.4$, the numerical calculation shows the number of
e-folding of the tachyonic LQC is
\[
N_{quan}\left| _{\phi _i=0.001,\dot{\phi}_i=0.0282}\right.
=\int_{t_i}^{t_{end}}\sqrt{\frac \kappa 3\rho _\phi \left(
1-\frac{\rho _\phi }{\rho _c}\right) }dt\simeq 65
\]
where $V_0$ takes the value as in the above.

Furthermore, it is also interesting to analyze the slow roll
inflation for other choice of the potential. For example, for an
inverse square potential \cite{padman} one can compare the quantum
inflation with the classical case.

\section{Conclusion}

\label{Sec. 4}In this paper we discuss the tachyon field in the
context of LQC. LQC essentially incorporates the discrete quantum
geometry effect. So, in the high energy region (approaching the
critical density $\rho _c$), LQC greatly modifies the classical FRW
cosmology and predicts a nonsingular bounce at the critical density
$\rho _\phi =\rho _c$. We show that this is always true in the
tachyon matter LQC. For the tachyon matter LQC, a
superinflation can appear in the region $\frac 12\rho _c<\rho _\phi <\rho _c$%
. This superinflation phase purely comes from the quantum effect. In
FRW cosmology, the state parameter equation $\omega $ of the tachyon
field belongs to $\left[ -1,0\right] $, so the superinflation phase
is lacking. In order to closely examine the modification to the
classical FRW equation, it is helpful to identify the effective
density $\rho _{eff\text{ }}$ and the pressure $P_{eff}$ based on
the modified Friedmann equation. We find that the effective density
and the pressure still satisfy the energy conservative equation.
Furthermore, the modified Raychaudhuri equation described by the
effective density and the pressure implies that the inflation phase
can be extended to the region where the classical inflation stops.
This effect is notable only when the inflation ends in a high energy
region.

The next issue what we have investigated is the evolution of the
tachyon field with an exponential potential. Using the numerical
method, we have found that, as in FRW cosmology, the solutions of
the tachyon field still keep the attractor behavior in LQC, and we
have found that with the evolution of tachyon field all the quantum
and classical trajectories approach each other and merge. At the
high energy region (approaching the critical density), the classical
trajectories deviate from the quantum ones. Evolving the tachyon
field backward on the boundary, the tachyon matter cosmology is
bounced into an contracting branch. In the phase space the first
quadrant describes the expanding branch, and the fourth quadrant
expresses the contracting branch. So, the evolution picture of the
tachyon field in LQC is this: in the distant past, the field-being
in the contracting branch with a negative velocity $-1$-is
accelerated to climb up the potential hill; then the field is
bounced into an expanding universe with positive velocity rolling
down to the bottom of the potential. What is more, the maximum
potential can be attained when the rolling velocity of the tachyon
field is equal to zero. In the slow roll limit, the number of
quantum e-folds is smaller than it is in the classical case. For
tachyonic LQC a sufficient inflation favors a bigger constant $V_0$
in the exponential potential or small tachyon mass $\alpha $.

We analyze the evolution of tachyon field with the exponential
potential in the context of LQC, and obviously, any other choice of
potential can be investigated via the same way. In fact, one may
also study cosmological consequences of tachyon matter field in the
context of the string theory. But as Kofman and Linde emphasized in
Ref. \cite{problem}, we face some problems with tachyon field, such
as the difficulty to obtain the inflation and the failure of
reheating, etc. LQC can naturally predict an inflation phase which
is independent of the choice of a particular potential and extend
the physical phase space of the tachyon field to the fourth
quadrant. We hope that this work can shed light on some of the
problems of understanding the tachyon field.

\acknowledgments

We are grateful Z.-X. Liu for discussion. We thank J. Wang and T.-X.
Zhang for their help. The work was supported by the National Basic
Research Program of China (2003CB716302).

\end{document}